\documentclass{elsarticle}
\usepackage{epsfig}
\usepackage{graphicx}
\usepackage{amsmath}
\usepackage[export]{adjustbox}
\usepackage{bm}

\def\ee{\end{eqnarray*}}
\def\be{\begin{eqnarray*}}
\def\bee{\end{eqnarray}}
\def\bbe{\begin{eqnarray}}

  \def\R{\mathrm{Re}}
   \def\I{\mathrm{Im}}

   \def\n{\bm n}
    \def\e{\mathrm e}

    \def\p{\partial}
    \def\q{\bm q} 

    \def\i{\mathrm i}

\def\d{\mathrm{d}}
\def\panel{\mathit{panel}}

\def\D{D}

\def\pp{\bm p}
\title{Straightforward integration for  free surface Green function and  body  wave motions}

\author{Zhi-Min Chen}
\address{School  of Mathematics and Statistics, Shenzhen University,
 Shenzhen 518060,  China}
\date{}
\journal{EJM/BF}
\begin{document}

\begin{abstract}   An alternative manner  is provided   for  solving the classical linearised problem of  the radiation and diffraction of
regular water waves caused by oscillation of a floating body in deep water. It is shown that the singular wave integrals of the three-dimensional free surface Green function $G$ and its    gradient  $\nabla G$ can be regarded as  regular wave integrals and are  integrated directly. The method is validated by comparing with benchmark data for a floating or submerged body undergoing oscillatory wave motions. The comparison shows that the  evaluation  is sufficiently accurate  for practical purposes. As the  significance of  the method,  the  numerical approximation stability for the gradient $\nabla G$  is shown to be the same with that for $G$.

\end{abstract}
\maketitle


\section{Introduction}

The determination of wave induced forces  resulting from body wave motion  is  a fundamental problem in  hydrodynamics.  For the linear situation, the velocity potential of the fluid motion problem is a harmonic function  and  can be  represented as a solution of body boundary integral equation involving  the pulsating  free surface Green function. The equation can be solved numerically  by combining  panel method and suitable approximation of the pulsating free surface Green function or free surface sources distributed on the body surface \cite{Frank1967, Lee1989, Lee1996, LN,Nob18}.  Varieties of Rankine simple source methods are also available to solve the body wave motion problems \cite{D,Y1981,Cao,Man,FengChen1,FengChen2} by using  the dynamic and kinematic free surface boundary conditions rather than employing free surface Green functions.  For a radial symmetric body
undergoing oscillatory wave motion, its linear analytic solution can be approximated by  a single free surface source rather than the boundary integral of free surface sources continuously distributed on the body surface.  For a heaving or surging hemisphere, the velocity potential  solution  is   decomposed into a  free surface source located at the centre of the sphere and a wave-free potential, which is expanded in a series of Legendre polynomials and sinusoidal functions \cite{U1949, H1955,Hu1982}.  The unknown source strength and expansion coefficients  are determined by the boundary condition of the velocity potential on the hemisphere. This method also applies to the wave resistance problem \cite{Fa} of a travelling  spheroid in waves and  is available to the understanding of   a submerged sphere in waves  \cite{W1986,Wu,p2013}.

\begin{figure}[h]
\centering \vspace{0mm}
\includegraphics[width=.8\textwidth]{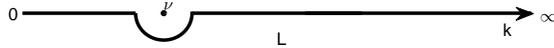}
\caption{Profile of the integration path $L$  in  (\ref{K3}) passing beneath the singular wave number $k=\nu$ in the complex plane. }
\label{f1}
\end{figure}

In the present study, we are interested in the approach of free surface Green function, which is evaluated in a straightforward manner.
Consider a fluid of
infinite water depth  upper bounded by the average  free water surface $z=0$ and consider a pulsating source   $\pp=(\xi,\eta,\zeta)$ with the unit  strength
undergoing periodic oscillatory motion with a constant frequency $\omega$ in the fluid.   The velocity potential of the source measured at a field point $\q=(x,y,z)$ is expressed as
 \be \Phi(\q) =\frac{1}{4\pi} \R(G(\q,\pp) \e^{-\i \omega t}).
 \ee
 Here $G$  is known as the fundamental solution of the Laplace equation under a free surface boundary condition and a radiation condition or  the pulsating free surface Green function
    \cite[pages 476-477]{W}
    \bbe \label{K1}
G\!\!\!&=\!\!\!&-\frac{1}{|\q-\pp|}-\frac{1}{|\q-\bar {\pp}|}-K(\q,\pp)
\bee
for the singular wave integral
\bbe K=
\frac {\nu}{\pi}\int^\pi_{-\pi}\int_L\frac{ \e^{k(z+\zeta)+\i k (x-\xi) \cos \theta+\i  k(y-\eta) \sin \theta} }{ k-\nu}
\d \theta \d k.\label{K3}
\bee
   Here $\nu = \frac{\omega^2}g$ with $g$ the gravitational acceleration and    $L$  illustrated by Figure \ref{f1} is an integration path passing  beneath  the singular wave number   $k=\nu$.

On the other hand, if we consider a three-dimensional body undergoing periodic oscillatory motion with a constant frequency $\omega$  in the fluid, the velocity potential of the linearised oscillatory fluid motion problem can be represented as
\bbe \Phi = \R(\phi \e^{-\i \omega t}), \label{new3}
\bee
where   $\phi$ is a stationary complex potential satisfying  the boundary integral equation
\bbe \label{bb}\phi(\q) +\frac1{4\pi}\int_{S}\phi(\pp)\n\cdot \nabla G(\q,\pp) \d S_{\pp}&=&\frac1{4\pi}\int_{S}G(\q,\pp)\n\cdot \nabla  \phi(\pp) \d S_{\pp}\\
&=&\frac1{4\pi}\int_{S}G(\q,\pp)(-\i\omega  n_\alpha) \d S_{\pp},\nonumber
\bee
after the use of  the impermeable body boundary condition
\bbe \n\cdot \nabla \phi = -\i\omega  n_\alpha\,\,\mbox{ on } S.\label{rad}
\bee
Here $\nabla =(\partial_\xi,\partial_\eta,\p_\zeta)$ is derivative operator with respect to $\pp$, $S$ is the average wetted surface of the body and $\n=\n(\pp)=(n_1,n_2,n_3)$ represents the normal vector field of $S$ pointing into the fluid.
 The body undergoes heave motion for $\alpha =3$, sway motion for $\alpha =2$ and surge motion for $\alpha =1$.

Equations (\ref{bb}) and (\ref{rad})  show that  the wave-body motion problem lies on the evaluation of the Green function  $G$ and its gradient $ \nabla G$ or the wave integral $K$ and its gradient $\nabla K$.

The present study is  a continuation of the author's previous examination  on the pulsating free surface Green function \cite{C2015} by integrating directly  a regular wave integral.
The method  \cite{C2015} is   based on the approximation
\bbe K=\lim_{\mu \to 0+}K^\mu \label{KK}
\bee
with respect to  the regular wave integral
\bbe
K^\mu\label{newKmu}
=
\frac {\nu+\i \mu  }{\pi}\int^\infty_0 \int^{\pi}_{-\pi}\frac{\e^{k(z+\zeta)+\i k (x-\xi) \cos \theta+\i k(y-\eta) \sin \theta }}{ k-\nu -\i  \mu }
\d \theta \d k.
\bee

As the normal derivative rather than the gradient is used in the boundary integral (\ref{bb}),  we   evaluate the normal derivative $\n\cdot \nabla G$ or $\n\cdot \nabla K$ instead of  the gradient $\nabla G$ or $\nabla K$.
 By the mathematical definition of the Riemann integrals  (\ref{newKmu}) and $\n\cdot \nabla K^\mu$, they can be  approximated respectively as integrals of   piecewise constant functions  within flat panels. However, one may consider higher order approximations to the integrals.  Integrating the regular wave integral (\ref{newKmu}) and the corresponding integral for $\n\cdot \nabla K^\mu$ straightforwardly, the evaluation is obtained as follows \cite{C2015}
\bbe
K \label{K}
&=&
\sum_{i=1}^{\infty}\sum_{j=1}^{\infty}c_{i,j} \e^{k_i(z+\zeta)+\i k_i (x-\xi) \cos \theta_j+\i k_i(y-\eta) \sin \theta_j },\label{KKmu}
\\
\n\cdot \nabla K&=& \label{DK}
\sum_{i=1}^{\infty}\sum_{j=1}^{\infty}c_{i,j}' k_i \e^{k_i(z+\zeta)+\i k_i (x-\xi) \cos \theta_j+\i k_i (y-\eta)\sin \theta_j} \label{nKmu1}.
\bee
with  respect to a small $\mu>0$, a set of mesh  grid points $\{ (k_i, \theta_j)|\, \, i=1, 2, ..., \infty; j=1, 2, ..., \infty\}$ of $[0,\infty)\times[-\pi,\,\pi]$ and  the expansion coefficients
\bbe
c_{i,j} &=& \frac{\nu+\i \mu}\pi (\theta_{j+1}-\theta_j) \ln \frac{k_{i+1} -\nu -\i \mu}{k_{i} -\nu-\i\mu},
\\
c_{i,j}'&=&c_{i,j}[n_3 -\i n_1 \cos \theta_j-\i  n_2 \sin \theta_j ].
\bee
This expansion gives rise to a simple evaluation of the Green function and, what is more, shows the structure  of a free surface wave (see the free surface wave elevation produced by the source in  \cite[page 174]{C2015}. That is, the singular wave integral is the superposition  of all incident wave potentials
 \bbe \label{wave}\e^{k_i(z+\zeta)+\i k_i (x-\xi) \cos \theta_j+\i k_i(y-\eta) \sin \theta_j }
 \bee
  in all incident angles   $-\pi \leq \theta_j <\pi$ and  wave numbers $0<k_i <\infty$.

However, it should be noted that the expansion (\ref{DK}) for the normal derivative of $\n\cdot \nabla K$ increases with the factor  $k_i$. Thus the convergence stability of   the $\n\cdot \nabla K$  expansion is very different to that of the $K$ expansion with respect to the free water surface $z+\zeta=0$.

 It is the purpose of the present paper to introduce  new expansion approximations of $K$ and $\n \cdot \nabla K$ so that the factor $k_i$ is removed and thus $\n \cdot \nabla K$ and $K$ have the same convergent property.

The singular wave integral  (\ref{KK})  has been approximated by  a variety of elementary function expansions (see, for example, \cite{N1985,Nob82,HW,24N,Nob17}).

The use of the regular wave integral (\ref{newKmu}) dates back to the work of Havelock \cite{H28,H32} in 1920s on   the regular wave integral
\bbe \label{m1m}\frac{\nu}{\pi} \R \int^{\pi}_{-\pi} \int^\infty_0 \frac{ \e^{k[z+\zeta + \i \cos (x-\xi)\cos \theta +\i (y-\eta)\sin\theta ]}}{k\cos^2\theta -\nu -\i \mu \cos \theta}dk d\theta \bee
 for  the translating  free surface Green function  with respect to  an artificial viscosity number $\mu$. The  regular wave integral
\bbe
\frac {2\nu(1+\i \mu)^2}{\pi}\int^\pi_{0}\int^\infty_0\frac{ \e^{k(z+\zeta)-\i k\sqrt{ (x-\xi)^2 + (y-\eta)^2} \cos \theta} }{ k-\nu(1+\i \mu)^2}
\d k \d \theta.\label{K33}
\bee
was also introduced  in  \cite{Nob82} to aid the evaluation of the pulsating free surface Green function in terms of exponential integrals  via  the  decomposition of  the singular wave integral into  a near field flow component defined by the singular wave number $k=\nu$ or the dispersion relation $k-\nu=0$ and a far field flow component defined by the integral away from $k=\nu$.  Further developments of this technique were  obtained in  \cite{Nob1992,Nob1995,Nob1996} for analytic derivations of a non-oscillatory near field flow component $G^N$  (or  $\phi^N$)  and a far field wave component $G^W$ (or  $\phi^N$) for the wave integral $G^F$ of a translating and pulsating free surface Green function (or   a free surface effect component   $\phi^F$ of a potential flow for a ship advancing in waves).
The far field wave components are single integrals resulted from  the integral along the domain defined by the poles of singular integral integrands. One may also refer to  \cite{N1987} on the translating Green function  for the use of (\ref{m1m}) for determining  the uniqueness of the corresponding singular integral. Recent development of \cite{Nob82} for polynomial function approximation to the two flow  components of the singular wave integral (\ref{KK}) was given in \cite{Nob17} and  applied to a body  wave motion problem \cite{Nob18}.

     The straightforward integration technique   is initiated  from  \cite{Chen2012}, on the integration of the regular wave integral of the two-dimensional vortex free surface Green function, which is developed from  the understanding of  the instability of viscous flow in  magnetohydrodynamics \cite{Chen2002,Chen2005}  dominated by the Harmann layer friction controlled by the Hartmann number $\mu$.  The limit (\ref{KK}) is used to assume $K=K^\mu$  for   $0<\mu\ll 1$ throughout our examination.   With the presence of  the parameter $\mu>0$, the real part of the integrand of (\ref{newKmu}) is smooth and symmetric around the wave  number $k=\nu$ and the corresponding imaginary part is smooth but close to  a dirac delta function.  Thus $K$ is integrable directly.
      In contrast to traditional evaluation schemes in  earlier examinations,     wave integral singularities   are   always a barrier in  Green function evaluations   as singular wave integrals
      are supposed  to be not  integrable directly.

  The pulsating free surface Green function applies to radiation and diffraction wave problem defined by the boundary integral equation (\ref{bb}) without involving integration  over the  water line, the intersection contour  of linear wetted body surface with average free water surface. However, if the body  advancing at a uniform speed, the water line integral $\int_\Gamma (G \p_x\phi -\phi\p_x G)dl$ arises due to  the integration by parts over free surface \cite{Bra, G}. Recently, a consistent boundary  integrating formulation  for  ship advancing in calm water  was  given in \cite{Nob2013,Huang} showing  that the troublesome water line integral of the function $G\p_x\phi$ can be cancelled with the boundary integration  over  the defference between the linearised wetted body surface and the averaging wetted body surface. 

\section{Evaluation of the Green function}

 With the use of the Bessel function \cite{Hand} of the first kind
 \be J_0(kR) &=&\frac{1}{2\pi} \int^{\pi}_{-\pi}
e^{\i k (x-\xi) \cos \theta + \i k(y-\eta) \sin \theta }d\theta
 = \frac{1}{2\pi} \int^{\pi}_{-\pi}
e^{\i k R \sin\theta  }d\theta
\ee
for $R=\sqrt{(x-\xi)^2+(y-\eta)^2}$,
the singular wave integral $K$ of the Green function  can be rewritten as  \cite[pages 476-477]{W}
\bbe K&=&\label{K0}
2\nu\int_L\frac{ \e^{k(z+\zeta)} }{ k-\nu}
J_0(kR) \d k.
\bee
Accordingly, the regular wave integral can be rewritten as
\bbe K^\mu&=&\label{K00x}
2(\nu+\i \mu)\int^\infty_0\frac{ \e^{k(z+\zeta)} }{ k-\nu-\i \mu}
J_0(kR) \d k.
\bee

In order to provide a convergent  evaluation of the gradient $\nabla K$ on the free surface $z+\zeta=0$, we use the derivative of the Bessel function of the first kind \cite{Hand}
\bbe \label{de}J_1(x)=-\frac{d J_0(x)}{dx}\bee
and the elementary identity
\bbe \frac{k}{k-\nu-\i\mu}= \frac{\nu+\i\mu}{k-\nu-\i\mu} +1\label{xy1}
\bee
to obtain
\begin{align}
\frac{\p K^\mu}{\p R}&=-2(\nu+\i \mu)\int^\infty_0  \frac{e^{k(z+\zeta )}k}{k-\nu-\i\mu}J_1(kR) d k\nonumber
\\
&=-2(\nu+\i \mu)^2\!\!\!\int^\infty_0 \!\!\! \frac{e^{k(z\!+\! \zeta )}}{k\!-\! \nu\!-\! \i\mu}J_1(kR) d k\!-\! 2(\nu\!+\! \i \mu)\int^\infty_0\!\!\! e^{k(z\!+\! \zeta )}J_1(kR) d k.\label{xy}
\end{align}
With the use of the derivative (\ref{de}) and integration by parts, the second integral on the right-hand side of the previous equation can be modified as, for $z+\zeta <0$,
\begin{align}
\int^\infty_0 e^{k(z+\zeta )}J_1(kR) d k
&=-\frac{1}{R}\int^\infty_0 e^{k(z+\zeta )}dJ_0(kR)\nonumber
\\
&=-\frac{1}{R} \left[e^{k(z+\zeta )}J_0(kR)\right]^{kR=\infty}_{kR=0}+\frac{1}{R}\int^\infty_0\!\!\! J_0(kR) \frac{\p \e^{k(z+\zeta )}}{\p k} d k\nonumber
\\
&=\frac{1}{R}+\frac{z+\zeta }R\int^\infty_0 e^{k(z+\zeta )}J_0(kR) d k\nonumber
\\
&=\frac{1}{R}+\frac{1}{R}\frac{z+\zeta }{\sqrt{R^2+(z+\zeta )^2}}, \,\,\mbox{ since } z+\zeta<0,\nonumber
\\
&=\frac{ R}{\sqrt{R^2+(z+\zeta )^2}(\sqrt{R^2+(z+\zeta )^2}+|z+\zeta |)},\label{xyz}
\end{align}
where we have used the identity
\bbe \frac1{\sqrt{R^2+(z+\zeta)^2}}=\int^\infty_0 e^{k(z+\zeta )}J_0(kR) d k
\bee
due to two-dimensional Fourier transform on the $Oxy$ plane. Therefore, the combination of (\ref{xy}) and (\ref{xyz}) gives
\bbe
\frac{\p K^\mu}{\p R}
&=&-2(\nu+\i \mu)^2\int^\infty_0  \frac{e^{k(z+\zeta )}}{k-\nu-\i\mu}J_1(kR) d k\nonumber
\\
&&-\frac{2(\nu+\i \mu) R}{\sqrt{R^2+(z+\zeta )^2}(\sqrt{R^2+(z+\zeta )^2}+|z+\zeta |)}.\label{KRR}
\bee
By (\ref{K00x}) and (\ref{xy1}), we have
\bbe
\frac{\partial K^\mu}{\p z}&=&2(\nu+\i \mu)\int^\infty_0  \frac{e^{k(z+\zeta )}k}{k-\nu-\i\mu}J_0(kR) d k\nonumber
\\
&=&2(\nu+\i \mu)^2\int^\infty_0  \frac{e^{k(z+\zeta )}}{k-\nu-\i\mu}J_0(kR) d k+2(\nu+\i \mu)\int^\infty_0 e^{k(z+\zeta )}J_0(kR) d k\nonumber
\\
&=&(\nu+\i \mu) K^\mu+\frac{2(\nu+\i \mu)}{\sqrt{R^2+(z+\zeta )^2}}.\label{Kz}
\bee
Let us note that the   identity $\p K/\p z=\nu K + 2\nu /\sqrt{R^2+(z+\xi)^2}$ is implied from  \cite[Eq. (9.3)]{Nob82}.

On the other hand, for the regular wave integral $K^\mu$ , we use the Bessel function asymptotic behaviors
\bbe
J_0(s) = O (\frac 1{\sqrt{s}}) 
 \,\, \mbox{ and }\,\, J_1(s) = O (\frac 1{\sqrt{s}}) 
\bee
with respect to large $s>0$ to obtain the convergence of  the infinite domain integral
\bbe
\int^\infty_0\frac{ \e^{k(z+\zeta)}J_n(kR) }{ k-\nu-\i \mu}
 \d k= \lim_{N\to \infty}\int^N_0\frac{ \e^{k(z+\zeta)} J_n(kR)}{ k-\nu-\i \mu}
 \d k,\,\,\, n=0, 1.
\bee
Therefore, we may define  $k_{\mathrm{max}}$  as  a large number $N$.
For a  dense coordinate  grid $\{ k_j\}_{j=1}^{N_k}$ of the interval $[0, \, k_{\mathrm{max}}]$, we have
\bbe K^\mu&=&2(\nu+\i\mu)\int^{k_{\mathrm{max}}}_0\frac{ \e^{k(z+\zeta)}J_0(kR) }{ k-\nu-\i\mu}\d k\nonumber
\\
&=&2(\nu+\i\mu)\sum_{j=1}^{N_k}\int^{k_{j+1}}_{k_j}\frac{ \e^{k(z+\zeta)}J_0(kR) }{ k-\nu-\i\mu}\d k.
\nonumber
\bee
This yields, by the continuous function property of the integrand numerator,
\bbe
K^\mu &=&2(\nu+\i\mu)\sum_{j=1}^{N_k}\e^{k_j (z+\zeta)}J_0(k_j R)\int^{k_{j+1}}_{k_j}\frac{ \d k }{ k-\nu-\i\mu}\nonumber
\\
&=&2(\nu+\i\mu)\sum_{j=1}^{N_k}\e^{k_j (z+\zeta)}J_0(k_j R)\ln\frac{k_{j+1}-\nu-\i\mu}{k_j-\nu-\i\mu}.\label{K00}
\bee
This also evaluates  $\frac{\p K^\mu}{\p z}$ due to (\ref{Kz}).

\begin{figure}
 \centering
              \includegraphics[height=.41\textwidth, width=.51\textwidth]{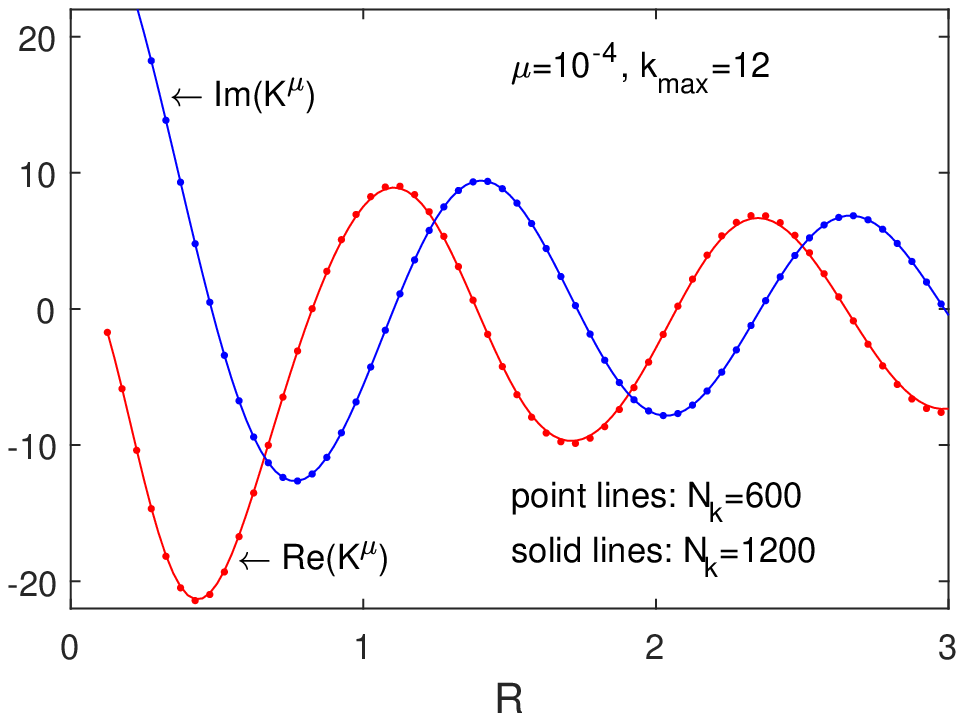}
 \hspace{-6mm}\includegraphics[height=.41\textwidth, width=.51\textwidth]{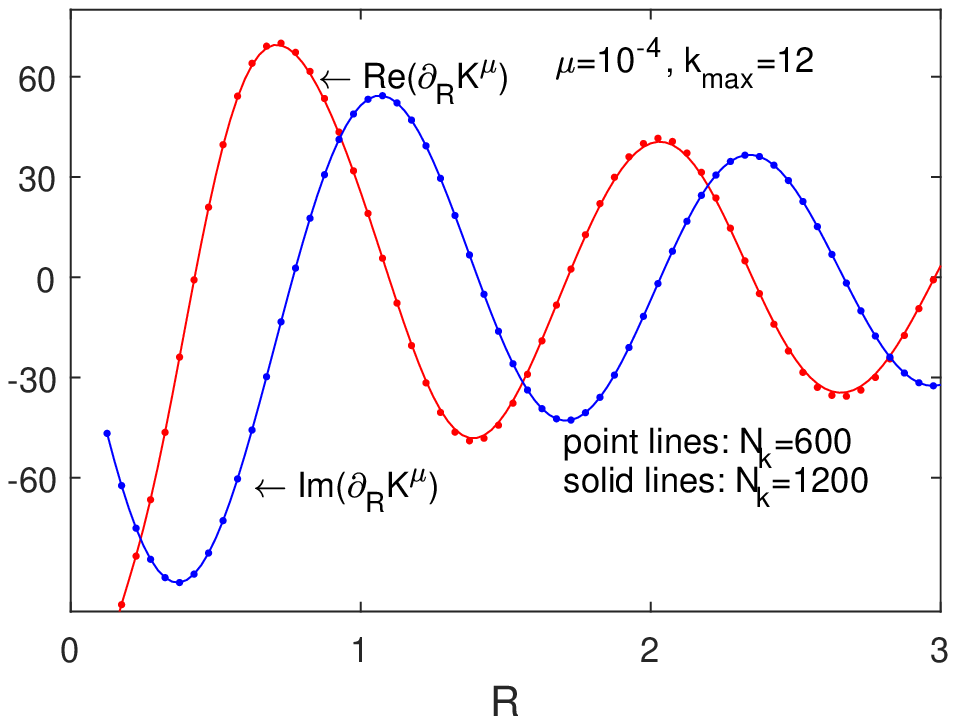}

               \includegraphics[height=.41\textwidth, width=.51\textwidth]{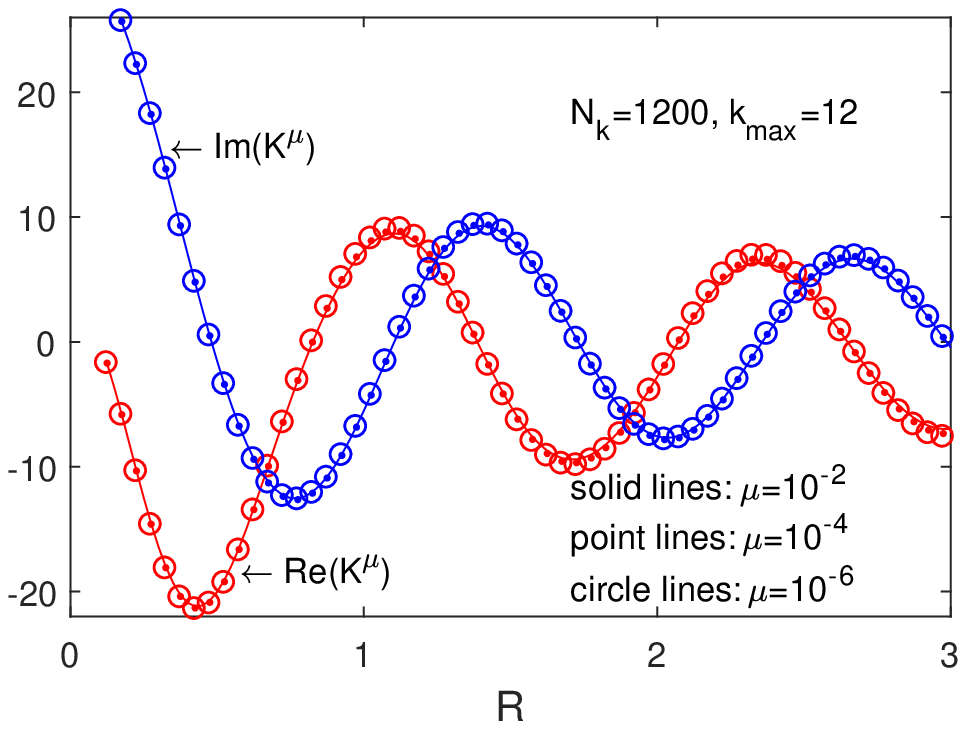}
 \hspace{-6mm}\includegraphics[height=.41\textwidth, width=.5\textwidth]{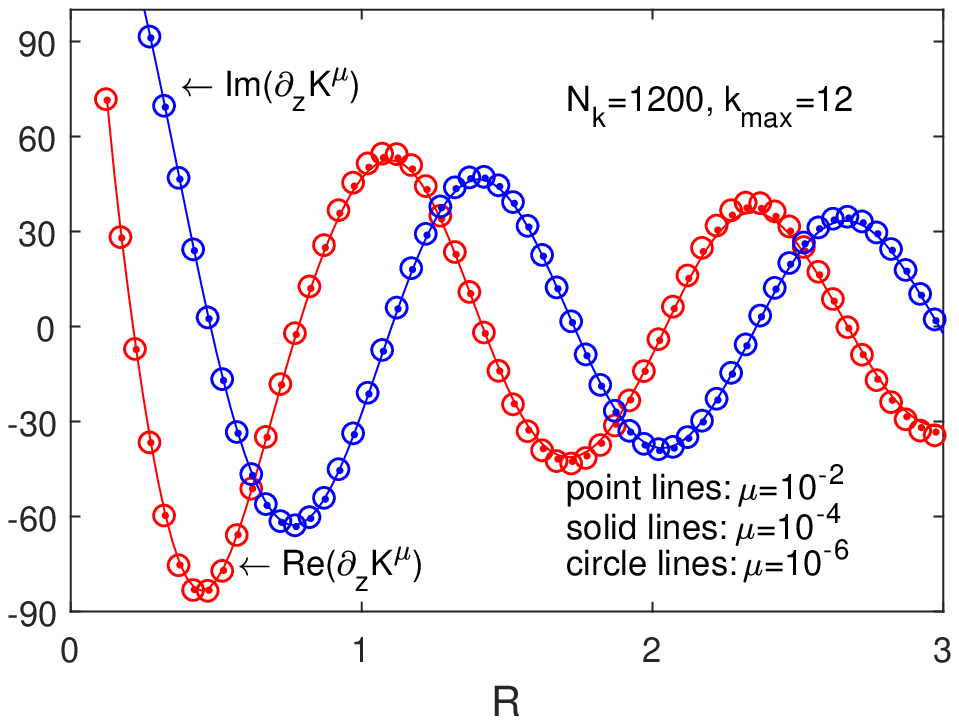}

               \includegraphics[height=.41\textwidth, width=.51\textwidth]{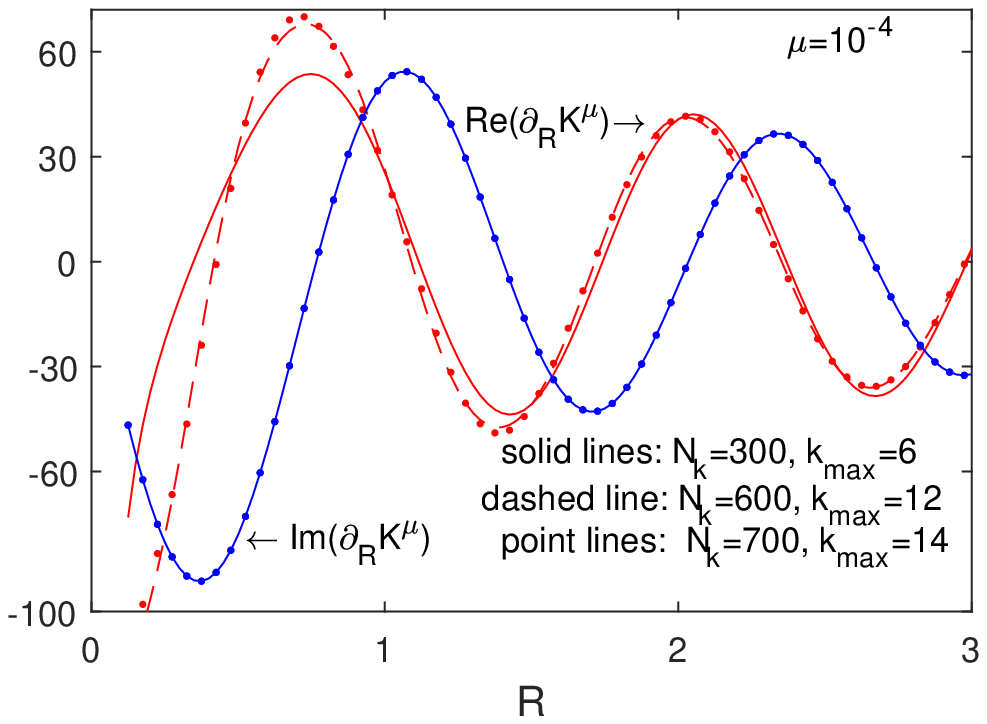}
 \hspace{-6mm}\includegraphics[height=.41\textwidth, width=.51\textwidth]{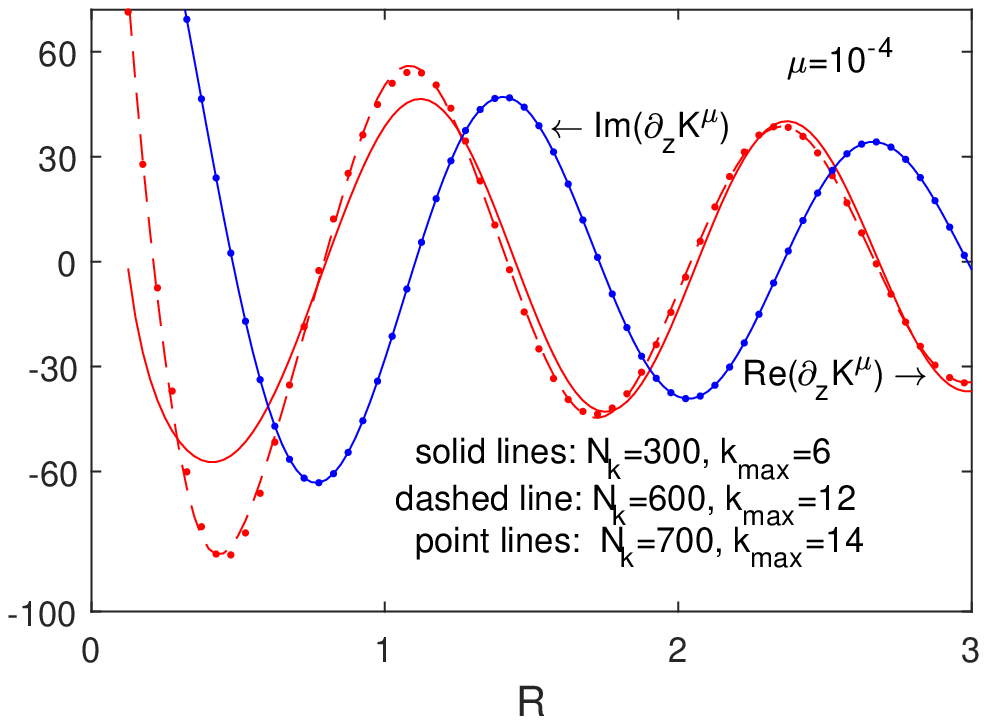}
 \caption{Convergence  of the regular wave integral $K^\mu$ in (\ref{K}) and its derivatives in (\ref{KR}) and (\ref{Kz}) with respect to different parameters at $z=0$ and $\nu=5$.}
 \label{f2}
 \end{figure}
Similarly, we have the partial derivative
\bbe\label{KK1}
\frac{\p K^\mu}{\p R}\nonumber
&=&-2(\nu+\i\mu)^2\sum_{j=1}^{N_k}\e^{k_j (z+\zeta)}J_1(k_j R)\ln\frac{k_{j+1}-\nu-\i\mu}{k_j-\nu-\i\mu}
\\
&&-\frac{2R(\nu+\i\mu)}{\sqrt{R^2+(z+\zeta )^2}(\sqrt{R^2+(z+\zeta )^2}+|z+\zeta |)}.\label{KR}
\bee

Upon the observation
\be \lim_{\mu \to 0+}K^\mu =K,\ee
we may assume $K=K^\mu$ for  small $\mu>0$ and thus have  the following approximations
\bbe 
K
&=&
\sum_{j=1}^{N_k}c_{j} \e^{k_j(z+\zeta) }J_0(k_j R),\label{Kmu}
\\
\frac{\p K}{\p R}
&=&-(\nu\!+\!\i\mu)\sum_{j=1}^{N_k}c_j\e^{k_j (z+\zeta)}J_1(k_j R)\nonumber
\\
&&
-\frac{2R(\nu\!+\!\i\mu)}{\sqrt{R^2\!+\!(z\!+\!\zeta )^2}(\sqrt{R^2\!+\!(z\!+\!\zeta )^2}\!+\!|z\!+\!\zeta |)},\label{KRmu}\\
\frac{\partial K}{\p z}
&=&(\nu+\i\mu)K+\frac{2(\nu+\i\mu)}{\sqrt{R^2+(z+\zeta )^2}}\label{Kzmu}
\bee 
 for
\bbe c_j =2(\nu+\i\mu)\ln\frac{k_{j+1}-\nu-\i\mu}{k_j-\nu-\i\mu}.
\bee

Moreover, we have  the normal derivative approximation
\bbe
\n\cdot \nabla K&=& \frac{\partial K}{\p R}\left(\frac{(\xi-x)n_1+(\eta-y)n_2}{R}\right)+n_3\frac{\partial K}{\p z}
\bee
or
\begin{align}
\hspace{-5mm}\frac{\n\cdot \nabla K}{\nu+\i\mu}
&=-\left(\frac{(\xi-x)n_1+(\eta-y)n_2}{R}\right)\sum_{j=1}^{N_k}c_j\e^{k_j (z+\zeta)}J_1(k_j R)+n_3K\nonumber
\\
&-\frac{2[(\xi-x)n_1+(\eta-y)n_2]}{\sqrt{R^2+(z+\zeta )^2}(\sqrt{R^2+(z+\zeta )^2}+|z+\zeta |)}+\frac{2n_3}{\sqrt{R^2+(z+\zeta )^2}}.\label{nK}
\end{align}

The validity of the approximation expansions (\ref{Kmu})-(\ref{Kzmu}) are essentially controlled by the quantities $\mu$, $N_k$ and $k_{\mathrm{max}}$. For displaying purpose, we take $z=0$ and $\nu=5$ to show the convergence of $K^\mu$ and its derivatives with respect to the approximations (\ref{Kz}), (\ref{K00}), (\ref{KR}) or (\ref{Kmu})-(\ref{Kzmu}) in Figure \ref{f2}.   It is shown in Figure \ref{f2} that $K^\mu$ and $\p_zK^\mu$ (and the same with $\p_R K^\mu$ actually) remain fixed for $0<\mu \leq  10^{-4}$. That is,  we may assume $K=K^\mu$ for $0<\mu \leq 10^{-4}$. Thus we usually take $\mu=10^{-4}$ in our computations. Moreover,  it is illustrated in Figure \ref{f2}  that we may take $N_k=600$ and $k_{\mathrm{max}}=12$ in the approximation expansions (\ref{Kmu})-(\ref{Kzmu}).

 By the linear dynamic free surface boundary condition, the free surface wave elevation $\chi$ produced by the single source $\pp=(\xi,\eta,\zeta)$ with the strength $4\pi$ can be derived as
 \bbe \chi  &=& -\frac1g \p_t \R ( G\e^{-i\omega t})
 = \frac\omega g  \R (\i  G\e^{-i\omega t})\nonumber
 \\
 &=&-\frac{\omega}{ g} \R\left([\frac2{\sqrt{R^2+\zeta^2}}+\sum_{j=1}^{N_k}c_{i} \e^{k_j\zeta }J_0(k_j R)]\e^{-\i \omega t}\right).
 \bee
That is, the surface wave produced by the singular source is  the superposition  of all radiation  waves
 $$\e^{k_j\zeta }J_0(k_j R)$$
  for wave numbers $0<k_j <\infty$.
 Therefore, velocity potential of a wave-body motion  is also the superposition  of the incident wave potentials  in a similar manner, as the corresponding wetted body surface  consists of all the free surface  sources distributed on the body surface.

 In contrast to the approximation of the wave integral by the expansion of the plan wave potentials (\ref{wave}) in all directions, the present approximation is the expansion of the radiation wave potentials $J_0(k_j R)$ centred at
the source point.

\begin{figure}
 \centering
 \includegraphics[height=.41\textwidth, width=.51\textwidth]{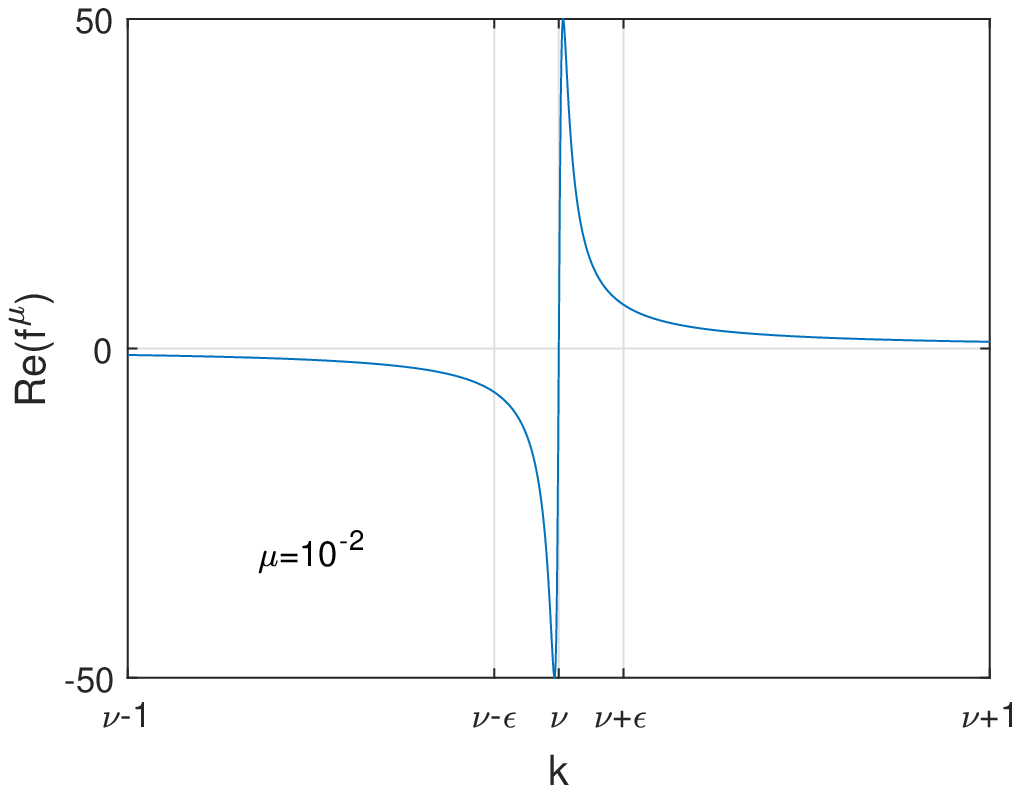}
  \hspace{-6mm}\includegraphics[height=.41\textwidth, width=.51\textwidth]{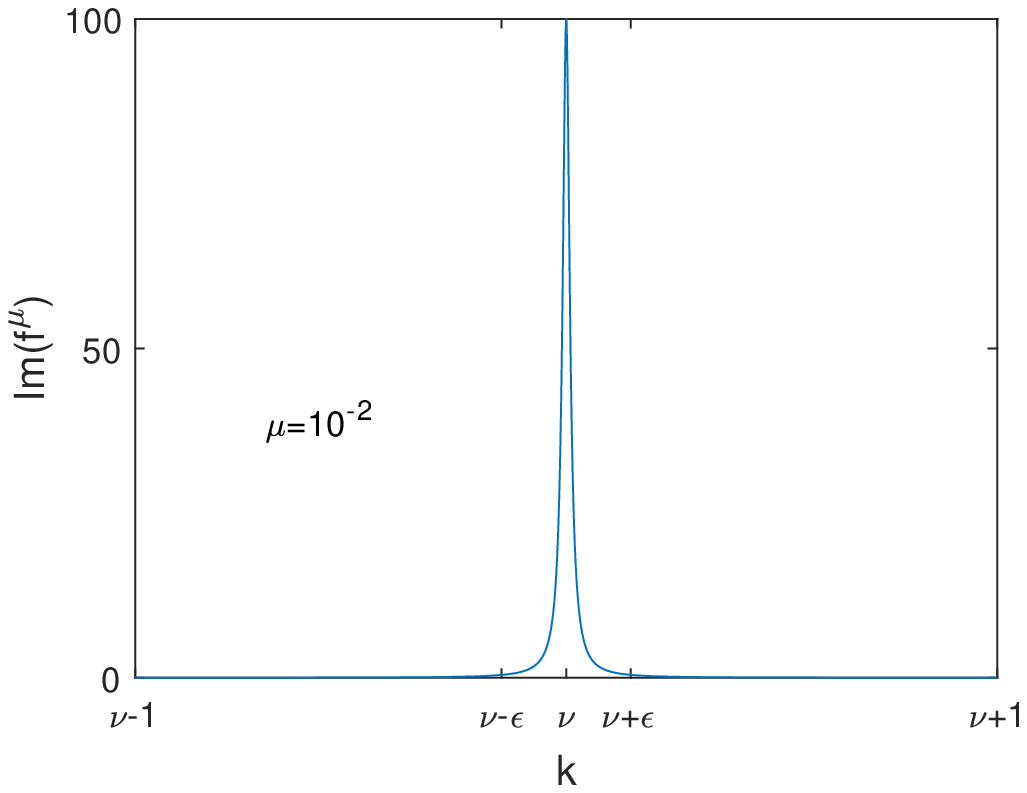}

 \includegraphics[height=.41\textwidth, width=.51\textwidth]{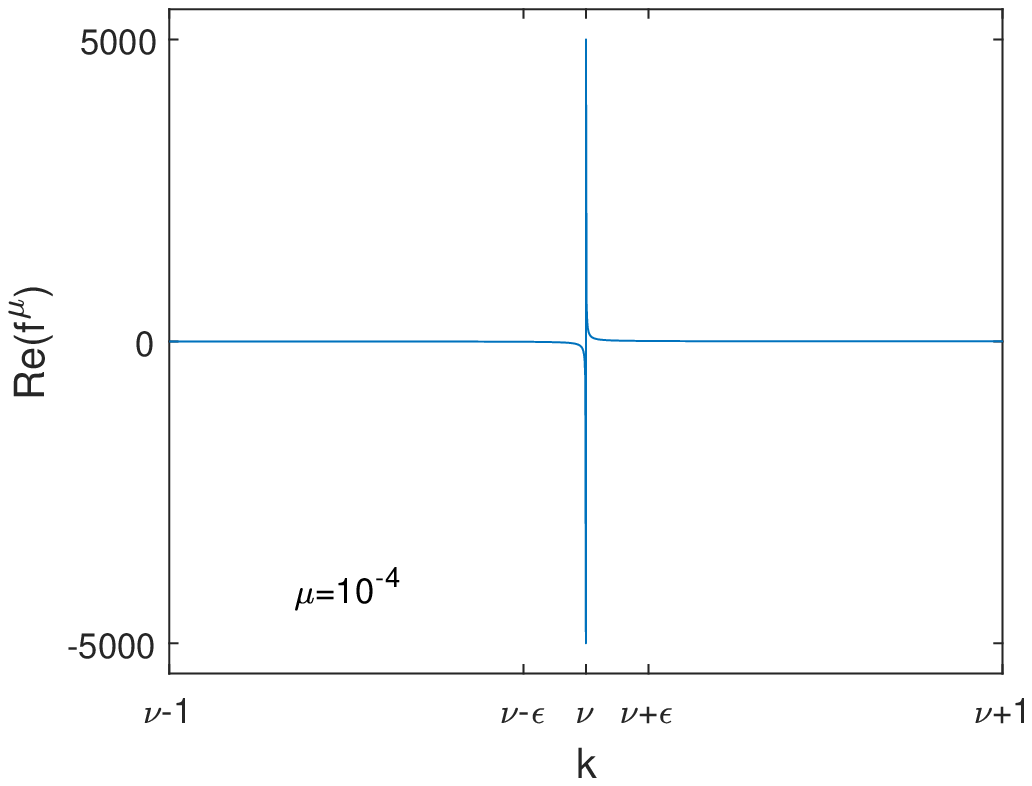}
  \hspace{-6mm}\includegraphics[height=.41\textwidth, width=.51\textwidth]{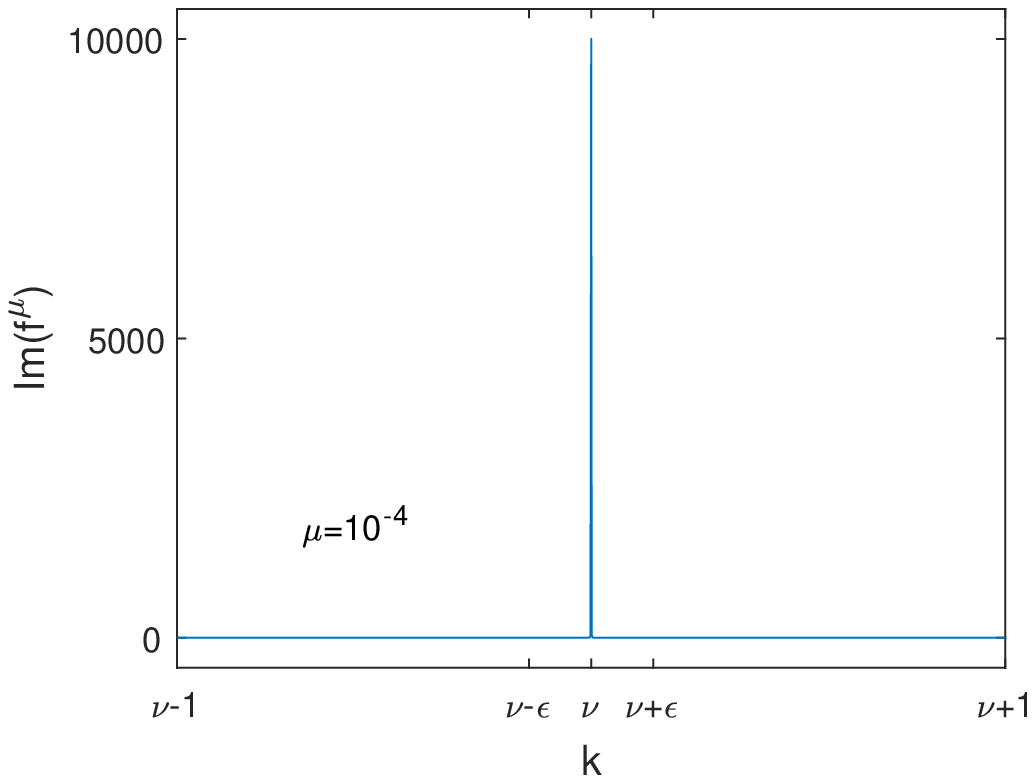}

 \includegraphics[height=.41\textwidth, width=.51\textwidth]{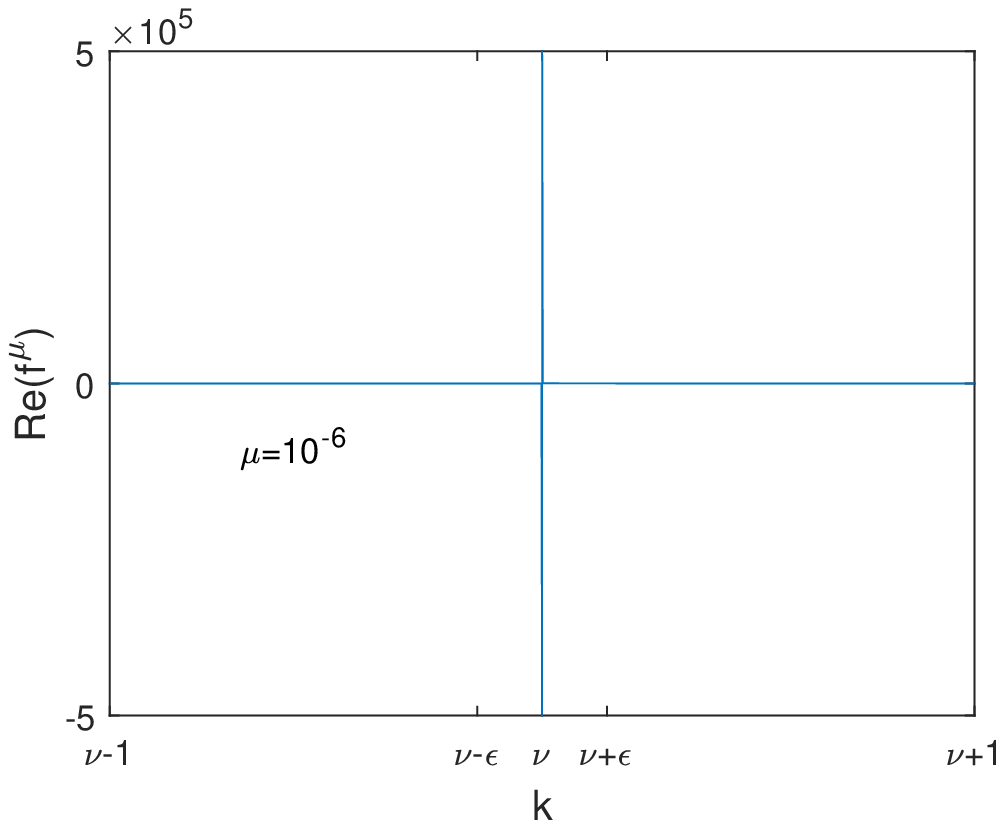}
  \hspace{-6mm}\includegraphics[height=.41\textwidth, width=.51\textwidth]{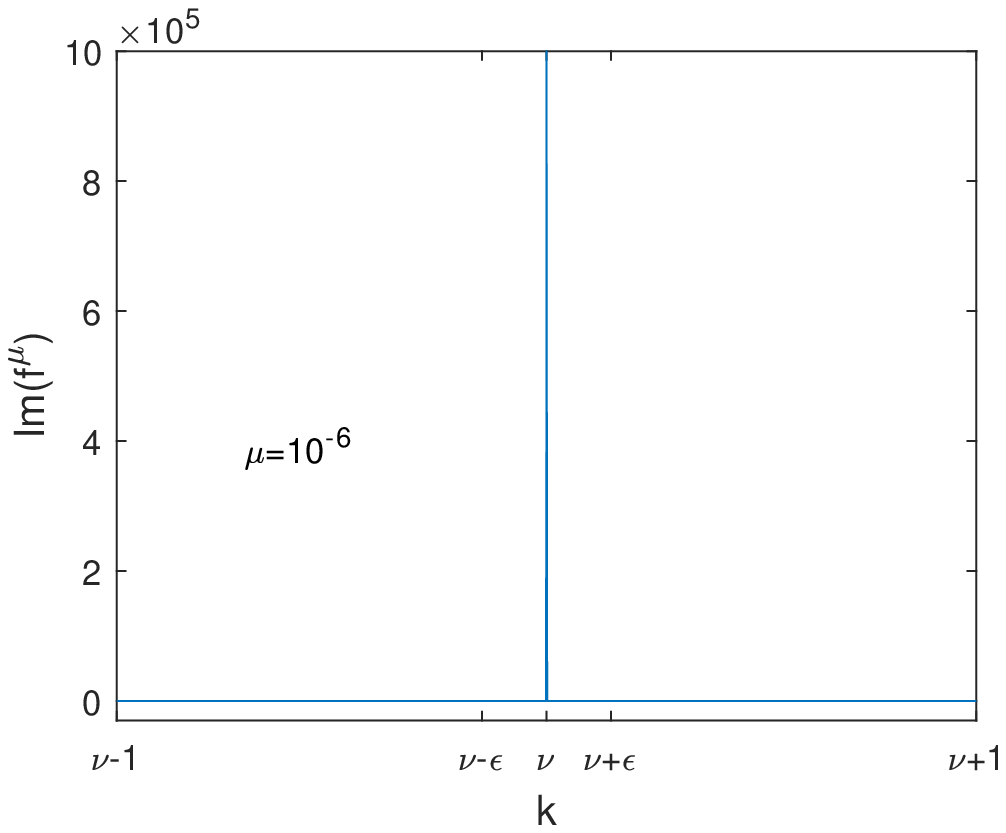}

 \caption{The analytical behaviour of the smooth integrand  $f^\mu=\frac1{k-\nu-\i\mu}$ for $\nu=3$ with respect to $0<\mu \ll \epsilon \ll  1$.}
 \label{f3}
 \end{figure}

  For convenience, the  mesh grid  can be simply defined   by the wave numbers  $$k_j = \frac{(j-1)k_{\mathrm max}}{N_k-1},
   $$
   although it is more economic to a use a mesh with sparse grid points away from $k=\mu$.
   To understand the nature of straightforward integration of $K^\mu$ and its stability with respect to  $\mu$ and mesh grid points $\{k_j\}$, we consider the convergence of the expansion
  (\ref{Kmu}), which is actually   controlled by the behaviour of the items from the panel integral around  the  wave number $k=\nu$. Note that $\e^{k(z+\zeta) }J_0(kR)$ is a smooth function of $k$ around $k=\nu$ and the grid panel $[k_{j_0}, k_{j_0+1}]$ containing  the wave number $k=\nu$ is   sufficiently small. Thus it remains to check the analytical behaviour of the panel integrals
   \be &&\int^{k_{j_0+1}}_{k_{j_0-1}} \frac{dk}{k-\nu-\i \mu}  \ \mbox{ if } \ k_{j_0}=\nu,
   \\
   &&\int^{k_{j_0+1}}_{k_{j_0}} \frac{dk}{k-\nu-\i \mu} \  \mbox{ if } \ k_{j_0}<\nu<k_{j_0+1}.
   \ee
   Without loss of generality, we may assume $\nu$ the central point of a grid panel  $[k_{i_0}, k_{i_0+1}]$, or $[\nu-\epsilon, \nu+\epsilon]$ for $\epsilon=\frac{k_{i_0+1}-k_{i_0}}2$.  If we use the original integral route around the singular point $k=\nu$ as given in Figure \ref{f1} of $K$, we have for singular integral of the lower half circle around $k=\nu$
\be
\int_{ |k-\nu|=\epsilon, \,\R (k)\leq 0} \frac{dk}{k-\nu}
&=&\int^{2\pi}_{\pi} \frac{\epsilon \e^{\i\theta} \i d\theta}{\epsilon \e^{\i\theta}}=\i \pi.
\ee
On the other hand, we have the regular  integral
\be  \int^{\nu+\epsilon}_{\nu-\epsilon} \frac{dk}{k-\nu-\i\mu}&=&\ln \frac{\epsilon-\i \mu}{-\epsilon-\i\mu}
 =
\i \pi +\ln (1+\frac{-2\i \mu}{\epsilon+\i\mu}).
\ee
That is,
\be
\int^{\nu+\epsilon}_{\nu-\epsilon} Re\frac{dk}{k-\nu-\i\mu} = \ln \frac{|\epsilon-\i\mu|}{|\epsilon+\i \mu|}\to 0, \,\mbox{ as } \mu\to 0
\ee
and
\be
\int^{\nu+\epsilon}_{\nu-\epsilon} \I\frac{dk}{k-\nu-\i\mu}= \arctan \frac  \epsilon\mu -\arctan \frac{-\epsilon}\mu \to \pi \,\mbox{ as } \mu\to 0.
\ee
 This is also illustrated in  Figure \ref{f3}, which shows  that  the real part of the function $\frac1{k-\nu-\i\mu}$ is almost symmetric about the point $k=\mu$  and hence the sum of the positive area in the first quadrant and the negative area in the third quadrant tends to zero, while
 the imaginary part  of the function $\frac1{k-\nu-\i\mu}$ tends to the dirac delta function as $\mu\to 0$.

\section{Evaluation of  the body wave motion problem}

Now we consider the body oscillating in the fluid domain $D$.
For    a field point  $\q=(x,y,z)\in\D$, the velocity potential $\phi=\phi(\q)$ has been expressed as a solution of the boundary integral equation  (\ref{bb}).
When  the field point tends to the body boundary $S$,  Eq. (\ref{bb}) reduces to  the boundary integral equation
\bbe \phi(\q) +\frac1{4\pi}  \lim_{\q' \in \D, \q'\to \q}  \int_{S}  \phi(\pp)\n\cdot \nabla G(\q',\pp) \d  S_{\pp} = \frac{1}{4\pi} \int_{S} G(\q,\pp) \n\cdot \nabla \phi(\pp) \d S_{\pp} \label{boundary}\bee
for $\q\in S$.

Following the discretisation scheme of \cite{C2015},   the boundary integral equation (\ref{boundary}) is now evaluated by   the approximation expansions (\ref{Kmu}) and (\ref{nK}).  The Hess-Smith Rankine  formulation \cite{C2015,HS,HS2,Newman1986} is employed to calculate the panel integral of the Rankine source potential and its normal derivative.
To do so, the two-dimensional body surface $S$  is approximated
by a set of  mesh grid points $\pp_{i,j}$ with  $i=1,...,N+1$ and $j=1,..., M+1$  for  suitable integers $N$ and $M$.
A single  $\panel_{i,j}$ associated with a centre panel point $\q_{i,j}$ and a panel normal vector $\n_{i,j}$   pointing into the fluid domain  is  determined  by the  four vertices
$ \pp_{i,j},\,\pp_{i,j+1},\,\pp_{i+1,j+1}$ and $\pp_{i+1,j}
.$

The boundary integral equation (\ref{boundary}) is  approximated
in the form of the algebraic equation system, for $ I=1, ..., N$ and $ J=1, ..., M,$
\bbe \sum_{i=1}^N\sum_{j=1}^M\left(\delta_{I,J,i,j} +  \mathcal{A}_{I,J,i,j}\right) \phi(\q_{i,j})= \sum_{i=1}^{N}\sum_{j=1}^M\n_{i,j}\cdot \nabla \phi(\q_{i,j})\mathcal{B}_{I,J,i,j} \,\label{phip}\bee
 for the Kronecker delta function  $\delta$ with $\delta_{I,J,i,j}=1$ whenever  $I=i$ and $J=j$. The influence coefficients of (\ref{phip}) are evaluated as
\begin{align}
 \mathcal{A}_{I,J,i,j}\nonumber
 =&-\lim_{\q\in\D, \q\to \q_{_{I,J}}}\frac{1}{4\pi}\int_{\panel_{i,j}}\,\n_{i,j}\cdot\nabla \frac1{|\q- \pp|}\d S_{\pp}
 \\&-\frac{1}{4\pi}\!\!\int_{\panel_{i,j}}\!\!\!\!\!\!\!\n_{i,j}\!\cdot\!\nabla \frac1{| \q_{_{I,J}}-\bar {\pp}|}d S_{\pp}
  -\frac{|\panel_{i,j}|}{4\pi}\n_{i,j}\!\cdot\!\nabla K(\q_{_{_{I,J}}}, \q_{i,j}),
\label{new1}
\\
\mathcal{B}_{I,J,i,j}&=-\int_{\panel_{i,j}}\frac{\frac1{|\q_{_{I,J}}\!-\!\pp|}+\frac1{|\q_{_{I,J}}\!-\!\bar\pp|}}{4\pi} \d S_{\pp} \!-\!\frac{|\panel_{i,j}|}{4\pi}K(\q_{_{I,J}},\q_{i,j}).
\label{new2}
\end{align}
 Here $|\panel_{i,j}|$ denotes the area of $\panel_{i,j}$.

The evaluation of  the panel integrals of (\ref{new1}) and (\ref{new2}) involving the Rankine source potential $\frac1{|\q_{_{I,J}}- \pp|}$ and its image $\frac1{|\q_{_{I,J}}- \bar\pp|}$
 is obtained by the Hess-Smith quadrilateral integral method and has been detailed in \cite{C2015}.

Thus the algebraic  equation (\ref{phip}) together with the boundary oscillation condition (\ref{rad})  can be solved from the Gaussian elimination scheme for the determination of the  unknown $\phi(\q_{i,j})$.

\section{Numerical results}\label{abc}
\def\RR{r}

For  the validation of the numerical approximation scheme, selected numerical results with respect to the body wave motion problem (\ref{new3})-(\ref{rad}) are displayed. Comparisons with the benchmark  data given by Wang \cite{W1986} and
 Hulme \cite{Hu1982}
 will  be presented with respect   a submerged sphere and  a floating hemisphere respectively. Let $\RR$ be the radius of the sphere or the hemisphere. The submergence of the sphere is defined by the parameter $h$, which measures the vertical distance between the  calm  water surface $z=0$ and the centre of the sphere. The mesh grid points   are given by the spherical coordinates
\be
\pp_{i,j}=  (\RR\sin \tau_i \cos \kappa_j,\, \RR \sin \tau_i\sin \kappa_j,\,  -h  +\RR\cos \tau_i),\ee
for $i=1,...,N+1$ and $j=1,...,M+1$.
Here  $- \pi \leq \kappa_j < \pi$,  $ 0\leq \tau_i \leq \pi$ for the submerged  sphere and $ \pi/2\leq \tau_i \leq \pi$ for the floating hemisphere. 

As illustrated in Figure \ref{f2}, the parameter  $\mu=10^{-4}$ is selected. The numerical computation is stable with respect to the choice of the parameters $k_{\mathrm max}$, $N_k$, $N$ and $M$. For the comparison purpose with respect to  added mass and damping coefficients, they are selected as  $k_{\mathrm max}\leq 14$, $N_k\leq 1100$, $N\leq 20$ and $M\leq 40$. Here we only provide upper bounds of the parameter. Actually, the bounds can be lowered and are  dependent on individual motions. For example, for  the heave motion of the sphere submerged at the water depth $h/\RR  =1.5$, satisfactory numerical results can be produced by taking  $k_{\mathrm max}=6$, $N_k=500$, $N=12$ and $M=14$. However, for the floating hemisphere, smaller mesh panels and larger truncation integral interval $[0,k_{\mathrm max}]$ for the regular wave integral are required as the stability of the wave integral
is reduced around the free surface $z=0$.

For the numerical velocity potential solution $\phi=\phi_\alpha$ ($\alpha=1, 2, 3$) of the boundary value problem (\ref{rad}) and (\ref{phip}),  the linear hydrodynamic  pressure  is expressed as
\be
p_\alpha=-\rho \frac{\p \Phi_\alpha}{\p t} = \omega\rho\R\left( \i \phi_\alpha \e^{-\i \omega t}\right)
\ee
for $\rho$ the fluid density.
This  defines the hydrodynamic wave force exerted on the average wetted body surface $S$:
$$F_{\alpha,\alpha}= \int_S p_\alpha n_\alpha dS$$
and the non-dimensional added mass and damping coefficients $A_{\alpha,\alpha}$ and $B_{\alpha,\alpha}$:
\bbe  A_{\alpha,\alpha}+\i B_{\alpha,\alpha} = \frac 1{\omega V } \int_{S}\i\phi_\alpha n_\alpha \d S\label{add}
= \frac 1{\omega V } \sum_{i=1}^N \sum_{j=1}^M \i\phi_\alpha(\q_{i,j}) n_{\alpha}(\q_{i,j})|\panel_{i,j}|.\bee
Here   $V$ is the volume of the moving body with the wetted body surface $S$.
Especially,   $V= \frac43\pi \RR ^3$  for the submerged sphere and   $V= \frac23\pi \RR ^3$  for the floating hemisphere.

\begin{figure}
 \centering
 \includegraphics[width=1.05\textwidth,left]{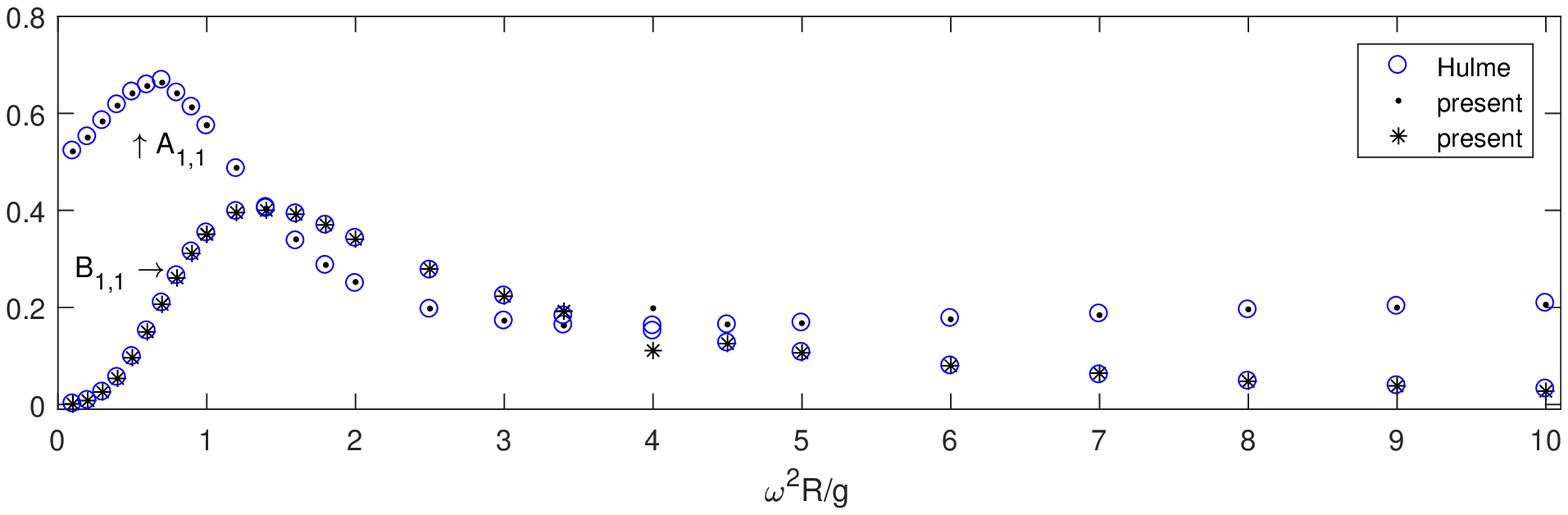}
\includegraphics[width=1.05\textwidth,left]{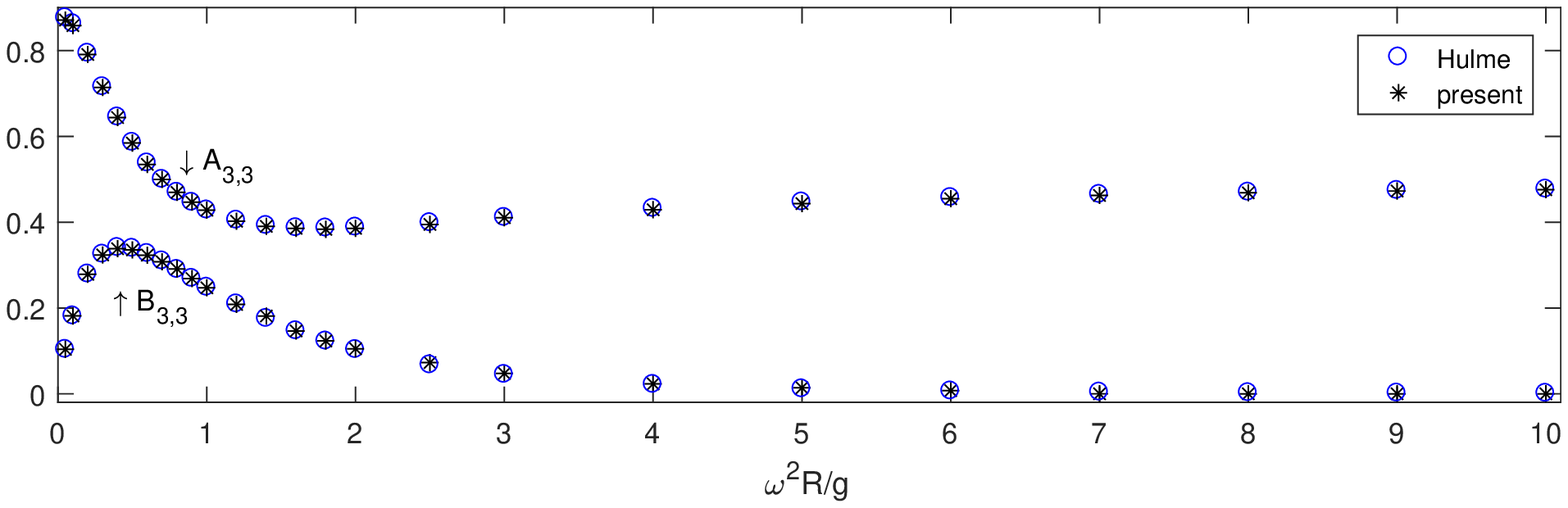}

 \caption{Added mass and damping coefficients produced by the present method and the semi-analytic solution data of Hulme \cite{Hu1982} for the    floating hemisphere in heaving or   surging motions.}
 \label{f4}
 \end{figure}
For the wave motion of the floating  hemisphere, present method  results  of  $A_{\alpha,\alpha}$ and $B_{\alpha,\alpha}$ for surging  motion ($\alpha=1$) and heaving motion ($\alpha=3$) are displayed in Figure \ref{f4}, which shows the existence of irregular frequency for $\omega^2\RR /g$ around  $4$ in a high frequency, the same phenomenon shown in \cite{C2015}.  As is well known \cite{Frank1967,Lee1989,U1981,Jon} that the combination of panel method and free surface Green function gives rise to  irregular frequencies in a high frequency range when a floating body undergoes oscillatory motions. Various  methods exist (see, for example, \cite{Lee1989,Lee1996,U1981,Lee1994}) to remove  non-physical irregular frequencies.  When $\omega^2\RR /g$ is away from the irregular frequency point, Figure \ref{f4}  presents excellent agreement between numerical solution and the semi-analytic solution of the celebrated  work  of Hulme \cite{Hu1982} for a heaving hemisphere.

\begin{figure}
\includegraphics[width=1.05\textwidth,left]{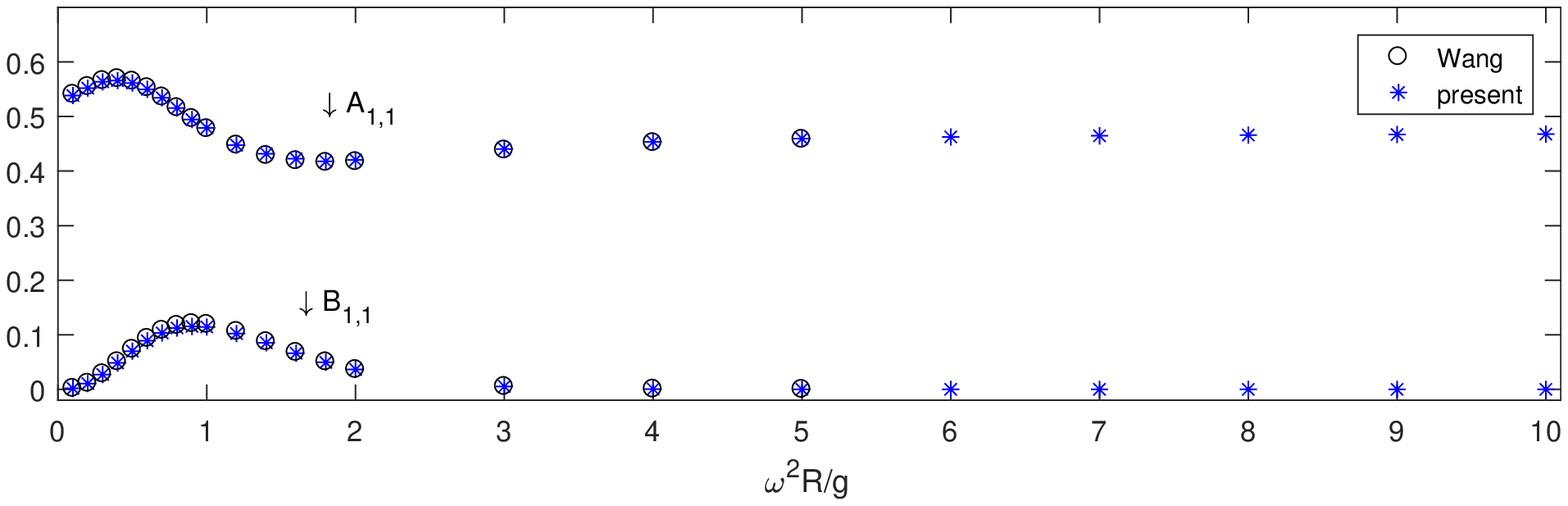}
\includegraphics[width=1.05\textwidth,left]{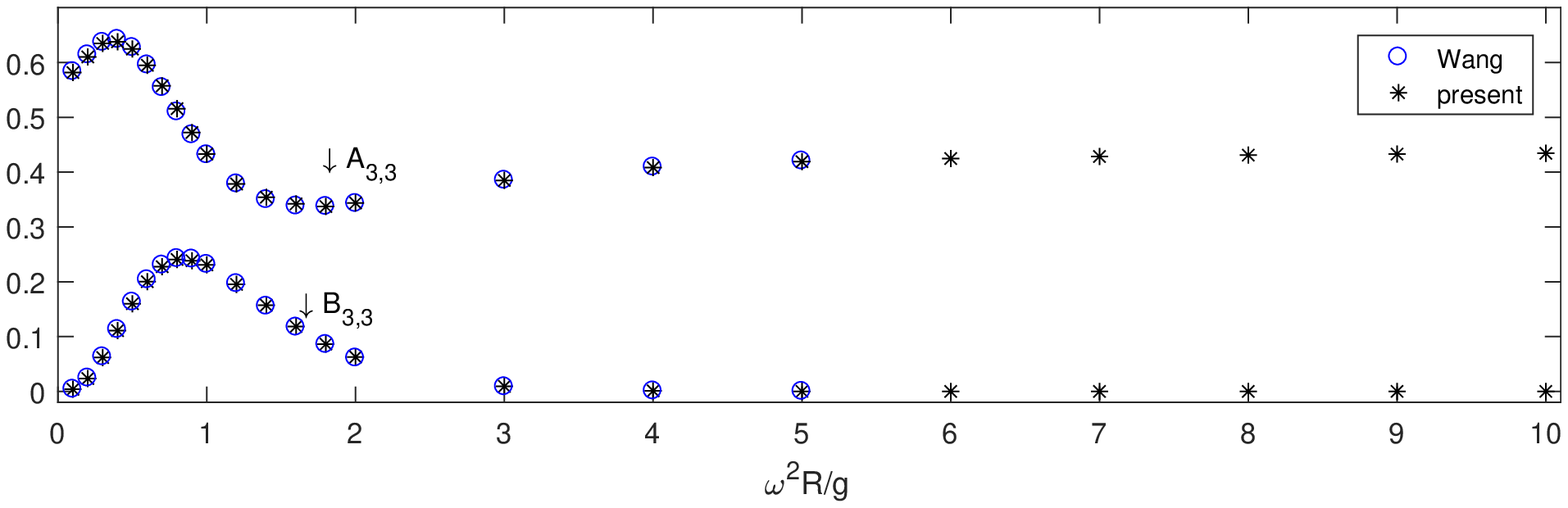}

 \caption{Added mass and damping coefficients produced by the present method and the semi-analytic solution data of Wang \cite{W1986} for the  heaving or   surging  sphere submerged at the water depth of $h/\RR =1.5$.}
 \label{f5}
 \end{figure}

For a fully submerged body of a radius $\RR $ at a submergence depth $h$, the irregular frequency phenomenon does not occur  due to stability improvement of the free surface Green function, since
the individual incident wave potentials
$ \e^{k_j(z+\zeta) }J_0(k_jR)
$
and
$ \e^{k_j(z+\zeta) }J_1(k_jR)
$
in (\ref{Kmu})-(\ref{Kzmu})
 are  controlled  by the exponentially decay function
$ \e^{k_j(z+\zeta)}$.
 Numerical solutions of heaving and surging motions with respect to the submerged sphere are presented in Figure \ref{f5} showing  that the present method solution agrees well with the semi-analytic  solution of  Wang \cite{W1986}.

\section{Discussion and conclusion} The free surface  Green function represents a pulsating  free surface source potential, which is the sum  of the Rankine simple Green function, its image with regarding to the average free water surface and a singular wave integral. Thence the evaluation of the Green function means that of the singular wave integral.

With the rapid development of computing capacity, numerical computation of a linear hydrodynamic problem  is no longer  time consuming. However, the Green function evaluation due to the presence of an singular wave integral is still known to be troublesome  and sophisticated mathematical treatments are supposed to be employed  to attack the singularity \cite{N1985,Nob82,HW,24N,Nob17}.
Therefore, the purpose of the present  investigation is not for  reducing  the numerical simulation time in  solving a linear hydrodynamics problem, but to simplify the  accessibility to coding a body wave motion flow.

   The problematic singular pole $k=\nu$ of singular wave integral   (\ref{K3})  is removable by using the  continuous  elementary function $\frac1{k-\nu-\i \mu}$  with  $0<\mu \ll 1$ in place of the unbounded function $\frac1{k-\nu}$ along the integration line $k>0$. Although the continuous function  reaches high peaks  $\pm \frac1{2\mu}$  for its real part and $\frac1\mu$ for its imaginary part in a vicinity of the wave number $k=\nu$, the combination of two real integration areas with respect to positive and negative  peaks $\pm \frac1{2\mu}$ around $k=\nu$  is zero, while the imaginary integration area with respect to the peak $\frac1\mu$ remains unchanged for small $\mu$ (see Figure \ref{f3}).  That is, the integration can be calculated  in the straightforward and simple way:
\bbe \int^{\nu+\epsilon}_{\nu-\epsilon} \frac{f(k)dk}{k-\nu -\i\mu}
=f(\nu)\int^{\nu+\epsilon}_{\nu-\epsilon} \frac{dk}{k-\nu -\i\mu}=f(\nu) \ln \frac{\epsilon-\i \mu}{-\epsilon -\i \mu}
= \i \pi f(\nu)
\bee
for a continuous function $f$ and parameters $0<\mu\ll \epsilon\ll 1$. Thus there is  no special treatment required with respect to the wave number  $k=\nu$.

  The author's previous study \cite{C2015}  shows that the direct integration of the wave integral to obtain the approximations   (\ref{K})-(\ref{DK}). In the present investigation, the expansions  (\ref{K})-(\ref{DK}) are simplified  to the form (\ref{Kmu})-(\ref{nK}). The approximation stability of  the gradient $\nabla K$ is significantly improved as the stability of the expansion for $\nabla K$ is the same with that for $K$ in  (\ref{Kmu})-(\ref{nK}).

  As a sample application, this scheme is employed to compute numerically the wave motion problems of a floating sphere and a submerged sphere respectively in harmonic waves  in order to compare with the benchmark data of \cite{Hu1982,W1986}. Figures \ref{f4} and \ref{f5} indicate the sufficient accuracy  to compute linear wave loads in practice.

The efficiency of the present scheme is twofold. Firstly,  we use the single integrals $K^\mu$, $\p_R K^\mu$ and $\p_z K^\mu$ rather the double integrals in \cite{C2015}.  Secondary,  the partial derivative $\p_z K^\mu$ in (\ref{Kz}) is in a linear form of $K^\mu$ in (\ref{K00x})  and,  moreover, the partial derivative  $\p_R K^\mu$  in (\ref{KRR}) is also   a linear form of $K^\mu$ if  $J_1$ is replaced by $J_0$. Thus the numerical computation of the gradient  $\nabla K^\mu$  essentially  becomes  that of $K^\mu$.

\

\noindent \textbf{Acknowledgement.} This work is  supported by NSFC of China (11571240).

\

\end{document}